\documentclass[twocolumn,prb,showpacs]{revtex4}
\usepackage{psfrag}
\usepackage{graphicx}
\usepackage{amsmath}
\usepackage{amssymb}

\begin{document}
\title{Towards a theoretical description of molecular junctions in the Coulomb blockade regime based on density functional theory}
\author{R. Stadler$^{1,2}$, V. Geskin$^1$ and J. Cornil$^1$}
\affiliation{$^{1}$Laboratory for Chemistry of Novel Materials, 
 University of Mons-Hainaut, Place du Parc 20, B-7000 Mons, Belgium\\ 
 $^{2}$Department of Physical Chemistry, University of Vienna
 Sensengasse 8/7, A-1090 Vienna, Austria}

\date{\today}

\begin{abstract}
Non-equilibrium Green’s function techniques (NEGF) combined with Density Functional Theory (DFT) calculations have become a standard tool for the description of electron transport through single molecule nano-junctions in the coherent tunneling regime. However, the applicability of these methods for transport in the Coulomb blockade (CB) regime is still under debate. We present here NEGF-DFT calculations performed on simple model systems in the presence of an effective gate potential. The results show that: i) the CB addition energies can be predicted with such an approach with reasonable accuracy; ii) neither the magnitude of the Kohn-Sham gap nor the lack of a derivative discontinuity in the exchange-correlation functional represent a problem for this purpose. 
 \end{abstract}
\maketitle

Interest in electron transport between nanoscale contacts has recently intensified, due to the advent of the technologically motivated field of molecular electronics and recent progress in experimental techniques for manipulating and contacting individual molecules~\cite{molelect}. Electron transport is generally described within two limiting regimes, namely, coherent transport (CT) for strong coupling between the molecule and the electrodes or Coulomb blockade (CB) for weak coupling. These two regimes can be considered as complementary in a quantum-mechanical sense~\cite{datta}: CT deals with evanescent waves passing through the junction and does not lead to a stepwise charging of the central molecule; in the CB (incoherent) regime, the electrons cross the junction one by one and get localized during a finite time over the molecule. Transport in the CB regime is characterized by so-called CB diamonds depicting frontiers between low- and high-conductivity domains in the source/drain bias – gate voltage coordinates~\cite{kouwen,kubatkin,nitzan}. Key quantities in this stability diagram are the energy differences between the ionization and affinity levels of a quantum dot or single molecule, which are referred to as addition energies E$_{add}$.

First-principle non-equilibrium Green's function (NEGF) methods are typically implemented~\cite{atk,xue,sanvito,kristian} in combination with density functional theory (DFT) to theoretically  describe electron transport through single molecule junctions. This approach has been used in numerous works and proves to be very useful for the description of the CT regime. On the other hand, for electron transfer in the CB regime, an integer charge is transferred between the molecule and the leads and results in a relaxation of the electronic structure of the central molecule. It has been recognized that this process cannot be suitably described with DFT or spin-restricted Hartree-Fock (RHF) based ground-state calculations~\cite{ratner} and that a multi-determinant configuration interaction (CI) scheme, or Fock space, should provide a general solution to this problem~\cite{datta}. However, it is difficult to apply such a CI treatment to the open system of the whole molecular junction, especially because the leads are usually described by mean-field band structure calculations. Recently, the CB regime was described within a NEGF approach in conjunction with spin-polarized hybrid DFT~\cite{palacios,palacios1} and Hartree-Fock~\cite{hartree} methods for junctions based on finite clusters containing partially filled degenerate orbitals. Recent quasiparticle calculations based on the GW approximation were found to not accurately represent the impact of local spin and charge fluctuations on CB~\cite{GW}. 
 
It has also been suggested that a standard DFT framework has inherent problems for describing electron transport in both the CB and CT regime~\cite{sanvito1}, due to the self-interaction (SI) of electrons~\cite{perdew} in a Kohn-Sham framework and the lack of a derivative discontinuity (DD)~\cite{perdew1} in the evolution of the eigenenergy of the highest occupied molecular orbital (HOMO) as a function of its occupancy, which can be fractional. In Ref.~\cite{sanvito1}, these two issues have been portrayed as intimately linked and a SI correction scheme has been devised as a remedy. Independently, optimized effective potentials and hybrid methods have been employed for reducing or removing SI from DFT-based electron transport calculations~\cite{yang}.

\begin{figure}
    \includegraphics[width=0.95\linewidth,angle=0]{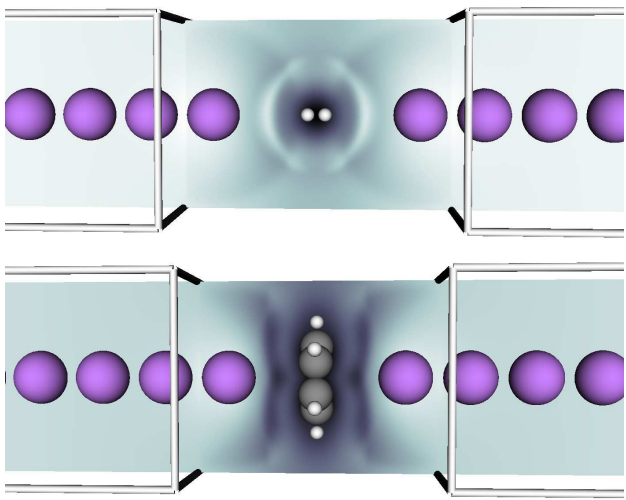}
      \caption[cap.Pt6]{\label{fig.strucpot} (Color online)Geometry and pattern of the applied gate potential in the two model molecular junctions of our study: a H$_2$ (top) and a benzene molecule (bottom) are attached to one-dimensional lithium electrodes, with a fixed separation of 4.37 and 4.00 \AA , respectively. The profile of the applied gate potential is shown with grey shading in both junctions, as obtained from the difference in the electrostatic potential between calculations performed with a V{\it{gate}} of 1V and 0V; the maximum of the potential is located in the black regions.}   
    \end{figure}

In this work, we present NEGF-DFT calculations performed in the presence of an effective gate potential V{\it{gate}}, which allows for an explicit charge transfer between one-dimensional lithium chains as electrodes, and H$_2$ and benzene molecules as the central unit (see Fig.~\ref{fig.strucpot}). The results obtained for these model systems show that: i) the size of the energy gap estimated from the electronic eigenenergies of the frontier Kohn Sham (KS) orbitals is not directly related to the threshold voltages or addition energies E$_{add}$ associated to the CB diamonds; ii) quantitatively realistic values for E$_{add}$ can be derived despite the fact that our calculations are carried out in the local density approximation (LDA) for the exchange-correlation functional (XC) and introduce a distortion in the charging process characteristic for DFT (in the sense that distinct one-electron transfer steps are replaced by continuous curves due the introduction of fractional charges N$_{add}$). 

Fig.~\ref{fig.strucpot} illustrates the two systems of our study, for which we performed NEGF-DFT calculations with the commercially available ATK software~\cite{atk1}. A single Li atom is included in the scattering region on each side of the central unit and each electrode part contains six atoms. A self-consistent solution for the electron density of the open system as a whole is carried out within the Keldysh formalism~\cite{atk,xue,sanvito,kristian} for every value of V{\it{gate}} on a grid of 0.5 V; a small source-drain bias of 20 mV has been applied in order to generate a current through the junction. The voltage V{\it{gate}} is introduced in the Keldysh Hamiltonian as $H_{\mu\nu}=H_{\mu\nu}+V{\it{gate}} S_{\mu\nu}$, where $S_{\mu\nu}$ is the overlap matrix and the indices $\mu$ and $\nu$ run only over the atomic basis functions of the molecules (i.e. V{\it{gate}} is not applied to the electrodes); the term containing V{\it{gate}} is added to the Hamiltonian at every step in the self-consistent cycle. The shape of the resulting potential is shown as gray shades in Fig.~\ref{fig.strucpot}. In all calculations presented in this article LDA has been used for the XC functional and a double-zeta polarized (DZP) basis set for both the molecules and the electrodes. 

  \begin{figure}   
  \includegraphics[width=0.95\linewidth,angle=0]{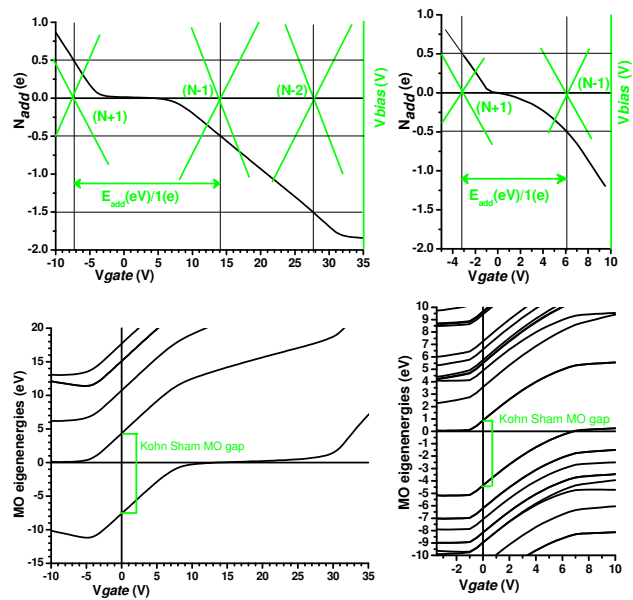}   
  \caption[cap.Pt6]{\label{fig.diamonds}(Color online)(Top) Evolution of the number of electrons on the molecules (N$_{add}$) as a function of V{\it{gate}} together with a schematic description of the CB diamonds at the predicted threshold voltages for the junction with H$_2$ (left) and benzene (right), respectively; (Bottom) The corresponding evolution of the MO eigenenergies as a function of V{\it{gate}}. N$_{add}$ has been determined by integrating charge density differences, with V{\it{gate}} = 0V taken as a reference and the spatial border between the molecule and the electrode defined as the region invariant in charge to V{\it{gate}}. The MO eigenenergies have been extracted from NEGF-DFT calculations by using a projection scheme~\cite{stokbro} and by defining the zero energy at the Fermi level of the Li electrodes. 
}    
  \end{figure}     

Fig.~\ref{fig.diamonds} shows the evolution of the charge on the molecules (N$_{add}$) and the KS-MO eigenenergies as a function of V{\it{gate}} for both systems. The range of gate voltages is larger in the positive region than in the negative one due to the well-known convergence problems for anions when using localized basis sets~\cite{jensen}. In contrast to the correct physical behaviour of electrons moving between the electrodes and the molecule one by one at given threshold voltages, N$_{add}$ evolves linearly from 0 to almost 2 electrons with Vgate due to the use of a closed-shell ansatz and the lack of a derivative discontinuity in DFT~\cite{datta,victor}. This general deficiency of DFT allows for any fractional value between 0 and 2 for the occupation of frontier orbitals of the molecules, even if they are only weakly coupled to metal electrodes (as it is the case in the present calculations).

Nevertheless, the addition energies Eadd associated to the CB diamonds can be calculated as:
\begin{eqnarray}
\label{eqn.central}
E_{add}=(E(N+1)-E(N))-(E(N)-E(N-1))= \nonumber \\
 \int_{0}^1N_{add}V{\it{gate}}(N_{add})-\int_{-1}^0N_{add}V{\it{gate}}(N_{add})= \nonumber \\
 V{\it{gate}}(N_{add}=+0.5e)-V{\it{gate}}(N_{add}=-0.5e)  \  \
\end{eqnarray}

In Eq.~\ref{eqn.central}, the first line is a general expression to calculate the difference between the redox levels from the total energies of the neutral and charged isolated molecules (as we did hereafter using Gaussian~\cite{gaussian}). In a second step threshold voltages obtained from NEGF-DFT calculations are introduced, where a midpoint rule for the integration over the charge has been used~\cite{williams}; E$_{add}$ is thus evaluated from the gate voltages required to add or subtract half an electron to the molecule. The second part of Eq.~\ref{eqn.central} is general in electrostatics and makes use of the relationship between a total energy E, a voltage V and a charge N~\cite{explain}. 

\begin{table}
\caption{\label{Eadd.tab}Addition energies E$_{add}$ calculated as V{\it{gate}} (N$_{add}$=+0.5e)-V{\it{gate}} (N$_{add}$=-0.5e) from NEGF-DFT for the junctions shown in Fig. 1, and as E(N+1)+E(N-1)–2 E(N) from the total energies of the neutral and charged isolated molecules. The KS HOMO-LUMO gap extracted from the NEGF-DFT calculations and reliable values for Eadd of the isolated molecules provided by experiment for H$_2$~\cite{jursic} and GW calculations for benzene~\cite{hybertsen} are also given for comparison. All values are given in eV.}
\vspace{0.5 cm}
{\centering
\begin{tabular}{c|c|c|c|c} \hline
Molecule & NEGF-DFT & DFT-E{\it{total}} & KS-MOs & Lit.(IP-EA)\\
\hline
H$_2$ & 21.30 & 21.35 & 11.95 & 18.56 \\
benzene & 9.80 & 11.54 & 5.23  & 10.51 \\
\hline
\end{tabular}\par}\end{table}

Table~\ref{Eadd.tab} collects the values of E$_{add}$ calculated for both junctions using the two expressions in Eq.~\ref{eqn.central}, with the same basis set and XC functional. For comparison, we also provide the Kohn-Sham HOMO-LUMO gap which has been derived from NEGF-DFT calculations at V{\it{gate}} = 0 V by projecting the eigenstates of the semi-infinite junction on the sub-Hamiltonian defined by the part of the basis set that is localized exclusively on the molecule~\cite{stokbro}; reliable values for E$_{add}$ are also listed for the isolated H$_2$~\cite{jursic} and benzene~\cite{hybertsen} molecules. N$_{add}$ has been estimated by integrating the voltage-dependent differences in the self-consistent charge density over the molecular part (see the caption of Fig.~\ref{fig.diamonds} for more details). The counter charges are located on the Li chains, with the main part on the atoms directly in contact with the molecule and the remainder decaying into the wires in Friedel oscillations (not shown here). We stress that the scaling factor $\alpha$ used in experiments~\cite{kouwen,kubatkin} to distinguish between the applied gate voltage and the effective potential felt by the molecule is equal to one in our calculations. 
Although the KS-MO gap is as expected about a factor of two smaller than all the other values, Table~\ref{Eadd.tab} shows that the results obtained from the NEGF-DFT approach match the two reliable sets of E$_{add}$ values obtained for the isolated molecules rather well.

\begin{figure}      
\includegraphics[width=0.95\linewidth,angle=0]{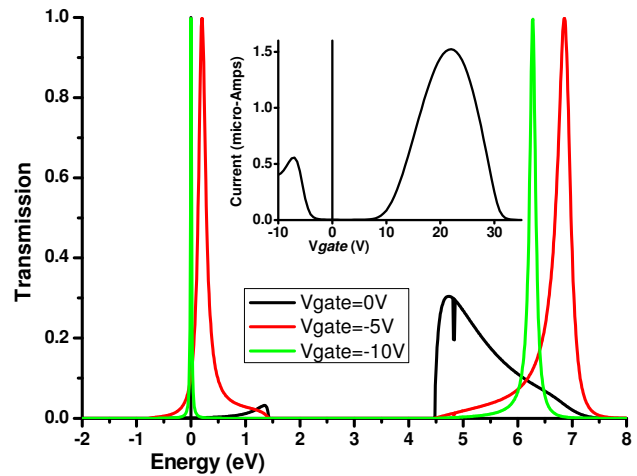}      
\caption[cap.Pt6]{\label{fig.trans} (Color online)Transmission functions for the Li-H$_2$-Li junction for different values of V{\it{gate}}, where the Fermi level E$_F$ of the Li chain defines zero on the energy axis. The peak corresponding to the LUMO moves from $\sim$+5 eV at V{\it{gate}} = 0V to $\sim$+0.2 eV at V{\it{gate}} = -5V; for V{\it{gate}} = -10V, it becomes pinned to E$_F$. In the transmission functions obtained for the two finite values of V{\it{gate}}, an additional peak associated to a higher-lying unoccupied MO can also be seen above 6 eV. The inset shows the current in the same junction with an applied source-drain bias of 20 mV as a function of V{\it{gate}}.  
}    
\end{figure} 

	The evolution of the MO eigenenergies as a function of the gate voltage in Fig.~\ref{fig.diamonds} reflect the corresponding curves obtained for N$_{add}$ since the occupation of a given MO can only vary between 0 and 2 while it stays pinned to the Fermi level. N$_{add}$ thus changes with V{\it{gate}} while the eigenenergies stay constant and vice versa. 
The pinning of the MOs to  Fermi level EF is also reflected in the transmission functions shown in Fig.~\ref{fig.trans} for the LUMO peak of the Li-H$_2$-Li junction. As long as the energy of the transmission peak does not reach E$_F$, it shifts as a function of V{\it{gate}} with a slope of one; this evolution is stopped when progressive charging is initiated in our mean-field calculations, instead of the correct stepwise charging, as already discussed above. As a consequence, the peak in the current / V{\it{gate}} curve shown in the inset of Fig.~\ref{fig.trans} is unphysically broadened by up to 10 V; in reality, this peak is expected to be much narrower due to the weak coupling between the molecule and the electrodes. The inset does show the high- and low-conductivity domains characteristic of the CB behavior when the gate voltage is modulated; the point is, however, that the position of the boundaries along the V{\it{gate}} axis is distorted by the mean-field character of the calculations and should be extracted properly in the way illustrated in Fig.~\ref{fig.diamonds} above. 

	It would be highly desirable to conduct similar calculations for more realistic electrodes than Li chains and for junctions where the molecule is chemisorbed on the leads via anchoring groups and the weakness of the coupling required for the CB regime promoted by alkyl spacers~\cite{kubatkin}. There are two main reasons preventing us to perform such calculations: i) the rather crude way used to introduce V{\it{gate}} by shifting the Keldysh Hamiltonian matrix elements in a defined region does not allow for the introduction of chemical bridges without ambiguity; and ii) due to the rather high voltages required for molecules with a moderate gap size and in view of the resulting significant perturbation of the electronic structure of the systems, the convergence of our calculations can only be achieved in a stepwise fashion, i.e. by using the converged density of a given voltage as a starting point for a next iteration, with steps of 0.5 V. Since this ultimately implies a large number of single self-consistent field calculations, replacing one-dimensional metallic chains by realistic surface electrodes is beyond our current computational resources. In this context, we cannot at this stage address the issue of screening or image charge effects which are highly relevant for the interpretation of recent experimental data. This is because such effects are expected to be only significant when considering large surface areas instead of surfaces consisting of just a single atom on each side. A recent comparison between DFT-LDA versus GW results for the adsorption of benzene on graphite~\cite{hybertsen} suggests that DFT is likely to fail in describing such polarization effects. We note, however, that the adsorbed molecule was not explicitly charged in this study and the focus was on the KS-MO gap, which is fundamentally different from our approach.

In conclusion, we have addressed the issue whether NEGF-DFT techniques, commonly used for the theoretical description of electron transport in the coherent tunneling regime, can also be employed for deriving relevant parameters in the Coulomb blockade regime. In our approach, we introduced explicitly a gate voltage in order to induce a charge transfer from the electrodes (Li chains) to the inserted molecules (H$_2$ and benzene). We obtained realistic results for addition energies (a key quantity in CB experiments) by using NEGF-DFT for the calculations on these simple model systems; we also argued that our results are not affected by the widely known short-comings of DFT such as the lack of a derivative discontinuity in the exchange-correlation functional. 

This research has been supported by the European projects SINGLE (FP7/2007-2013, no. 213609) and MODECOM (NMP3-CT-2006-016434), the Interuniversity Attraction Pole IAP 6/27 Program of the Belgian Federal Government « Functional supramolecular systems (FS2) », and the Belgian National Fund for Scientific Research (FNRS). We are especially indebted to Dr. J. A. Torres for his advice. J.C. is a Research Associate of FNRS. R.S. is currently supported by the Austrian Science Fund FWF, project Nr. P20267.

%%%%%%% References

\bibliographystyle{apsrev}

\begin{thebibliography}{23}
\expandafter\ifx\csname natexlab\endcsname\relax\def\natexlab#1{#1}\fi
\expandafter\ifx\csname bibnamefont\endcsname\relax
  \def\bibnamefont#1{#1}\fi
\expandafter\ifx\csname bibfnamefont\endcsname\relax
  \def\bibfnamefont#1{#1}\fi
\expandafter\ifx\csname citenamefont\endcsname\relax
  \def\citenamefont#1{#1}\fi
\expandafter\ifx\csname url\endcsname\relax
  \def\url#1{\texttt{#1}}\fi
\expandafter\ifx\csname urlprefix\endcsname\relax\def\urlprefix{URL }\fi
\providecommand{\bibinfo}[2]{#2}
\providecommand{\eprint}[2][]{\url{#2}}

\bibitem{molelect} Introducing Molecular Electronics. (Lect. Notes Phys. 680) G. Cuniberi, G. Fagas, K. Richter, Eds. Springer, 2005.
\bibitem{datta} S. Datta, cond-mat/0603034; B. Muralidharan, A. W. Ghosh, and S. Datta, Phys. Rev. B {\bf 73}, 155410 (2006); B. Muralidharan, A. W. Ghosh, and S. Datta, Molecular Simulation {\bf 32}, 751 (2006). 
\bibitem{kouwen}  L. P. Kouwenhoven, D. G. Austing, and S. Tarucha, Rep. Prog. Phys. {\bf 64}, 701 (2001). 
\bibitem{kubatkin}  S. Kubatkin, A. Danilov, M. Hjort, J. Cornil, J. L. Brédas, N. Stuhr-Hansen, P. Hedeg\aa rd, and T. Bj{\o}rnholm, Nature {\bf 425}, 698 (2003); A. Danilov, S. Kubatkin, S. Kafanov, P. Hedeg\aa rd, N. Stuhr-Hansen, K. Moth-Poulsen and T. Bj{\o}rnholm, Nano Lett. {\bf 8}, 1 (2008). 
\bibitem{nitzan}  A. Nitzan and A. Ratner, Science {\bf 300}, 1384 (2003). 
\bibitem{atk}  M. Brandbyge, J. L. Mozos, P. Ordejon, J. Taylor, and K. Stokbro, Phys. Rev. B {\bf 65}, 165401 (2002). 
\bibitem{xue}  Y. Xue, S. Datta, and M. A. Ratner, Chem. Phys. {\bf 281}, 151 (2002). 
\bibitem{sanvito}  A. R. Rocha, V. M. Garcia-Suarez, S. W. Baily, C. J. Lambert, J. Ferrer, and S. Sanvito, Nature Materials {\bf 4}, 335 (2005). 
\bibitem{kristian}  K. S. Thygesen and K. W. Jacobsen, Chem. Phys. {\bf 319}, 111 (2005). 
\bibitem{ratner}  S.M. Lindsay and M. A. Ratner, Adv. Mater. {\bf 19}, 23 (2007). 
\bibitem{palacios} J. J. Palacios, Phys. Rev. B {\bf 72}, 125424 (2005). 
\bibitem{palacios1}  I. Yanov, Y. Kholod, J. Leszczynski, and J. J. Palacios, Chem. Phys. Lett. {\bf 445}, 238 (2007). 
\bibitem{hartree} H. Wang and G. K.-C. Chan, Phys. Rev. B {\bf 76}, 193310 (2007).
\bibitem{GW} X. Wang, C. D. Spataru, M. S. Hybertsen and A. J. Millis, Phys. Rev. B {\bf 77}, 045119 (2008).
\bibitem{sanvito1} C. Toher, A. Filipetti, S. Sanvito, and K. Burke, Phys. Rev. Lett. {\bf 95}, 146402 (2005).
\bibitem{perdew} J. P. Perdew and A. Zunger, Phys. Rev. B {\bf 23}, 5048 (1981).
\bibitem{perdew1} J. P. Perdew, R. G. Parr, M. Levy, and J. L. Balduz, Phys. Rev. Lett. {\bf 49}, 1691 (1982).
\bibitem{yang} S. H. Ke, H. U. Baranger, and W. Yang, J. Chem. Phys. {\bf 126}, 201102 (2007).
\bibitem{atk1}  ATK, Version 2.2, atomistix a/s,  2004, www.atomistix.com.
\bibitem{stokbro} K. Stokbro, J. Taylor, M. Brandbyge, J. Am. Chem. Soc. {\bf 125}, 3674 (2003).
\bibitem{jensen} F. Jensen, J.Chem. Phys. {\bf 117}, 9234 (2002).
\bibitem{victor} V. Geskin, R. Stadler and J. Cornil, submitted (2008).
\bibitem{gaussian} M. J. Frisch, G. W. Trucks, H. B. Schlegel et al., Gaussian 98, Revision A.11.3, Gaussian, Inc., Pittsburgh PA, (2002); www.gaussian.com.
\bibitem{williams} A. R. Williams, R. A. deGroot and C. B. Sommers, J. Chem. Phys. {\bf 63}, 628 (1975).
\bibitem{explain}It might seem counterintuitive to rely on a threshold voltage for the transfer of half an electron since the concept of fractional charges is somewhat artificial in quantum theory. This concept is, however, implicit in DFT where the central variable is the electron density. In correlated quantum-chemical approaches, the relation between E, V, and N in Eq. (1) still holds since the transfer of  $\pm$0.5 e and $\pm$1.0 e occurs at the same voltages due to the step-like change in the charge N as a function of the voltage V.
\bibitem{jursic} B. S. Jursic, J. Mol. Struct. (Theochem) {\bf 491}, 11 (1999).
\bibitem{hybertsen} J. B. Neaton, M. S. Hybertsen and S. G. Louie, Phys. Rev. Lett. {\bf 97}, 216405 (2006).

\end{thebibliography}

\end{document}